\documentstyle[aps]{revtex}
\begin{document}
\draft
%
%
%
%
\title{
Exact Result on Hole Dynamics of \\
One-Dimensional Supersymmetric {\it t-J} Model\\
 with Long Range Interaction}
\author{Yusuke Kato\cite{permanent}}
\address{
Institut f\"ur Theoretische Physik, 
Universit\"at zu K\"oln, 
Z\"ulpicher str.77, 
D-50937, 
K\"oln, 
Germany}
\maketitle
\begin{abstract}
We obtain an exact result on single hole dynamics in one-dimensional Mott-insulators through the study on the supersymmetric {\it t-J} model with long range interaction at half-filling. We calculate the expression $G_{\rm 1s1h}(x,t) $ for  one spinon one holon contribution to the hole propagator. We find that $G_{\rm 1s1h}(x=0,t=0)=3/8$, which amounts to 75\% of the spin-fixed mean particle density $1/2$.
\end{abstract}
\pacs{75.10.Jm, 05.30.-d,71.27.+a}
%
%
%
%
%
%
%
Single hole motion in Mott insulators is a nontrivial problem in strongly correlated electron systems. This issue becomes particularly intriguing in some one-dimensional systems: a spin liquid ground state causes a novel effect on  hole motion; the underlying elementary excitations such as spinon (with spin 1/2 and no charge) and holon (charge e and no spin) provide a proper description to the hole dynamics. This situation seems to be realized in ${\rm SrCuO_2}$ \cite{Kim}; in the angle-resolved photo-emission experiment on that material, the authors of ref. \cite{Kim} identified dispersive peaks in the spectral intensity as holon and spinon dispersions. 
Meanwhile, the dynamics of the $U\rightarrow\infty$ Hubbard model has been studied intensively and important aspects of the spectral weight such as power-low singularity along the holon band \cite{SP} and the shadow band structure \cite{Penc2} have been clarified. However nontrivial is to what extent these properties survive in the {\it t-J} model with finite $J$. Further no simple analytical expression for the spectral weight has been obtained. Thus an analytical study on the hole dynamics in one-dimensional Mott insulators with finite $J$ would be worthwhile. 

As a solvable {\it t-J} model, we consider the one-dimensional supersymmetric {\it t-J} model with long range interaction\cite{KY}. This model has simple but nontrivial physical properties; spinon and holon form an ideal gas \cite{KatoKuramoto} obeying the fractional exclusion statistics \cite{Haldanefrac}; at half-filling, this {\it t-J} model reduces to the Haldane-Shastry (HS) model\cite{Haldane,Shastry}. The dynamical spin correlation function of HS model has been derived in an extremely simple form\cite{HZ}. Thus this model is eminently suitable for our purpose. 

For this {\it t-J} model, Ha and Haldane \cite{HaHaldane} studied the hole dynamics away from the half-filling and obtained the region of non-vanishing spectral weight of the spectral function in the frequency-momentum space (``compact support''). In addition, they identified excitation contents as simultaneous excitations of three spinons, three holons and one antiholon. At half-filling, however, the antiholons are suppressed. As a result, not only three spinon one holon (3s1h) state but also one spinon and one holon (1s1h) state contribute to the hole dynamics\cite{HaHaldane}. 

In this paper, we consider the spectral weight of the hole propagator, which has been left as an unsolved problem in\cite{HaHaldane}. To be precise, we calculate the exact expression for the (1s1h) contribution $G_{\rm 1s1h}(x,t)$ to the hole propagator at half-filling. Our calculation has two steps; first we map the {\it t-J} model into a multicomponent Calogero-Sutherland model \cite{Cal,Suth}. Next we calculate the propagator in the resulting model with the use of the theory of the nonsymmetric Jack polynomials  \cite{Opdam,Sahi} and their variants \cite{BF1,Dunkl}. In the following part, we first present our result and then explain the derivation.

We consider the {\it t-J} model        
\begin{equation}
{\cal H}_{tJ}={\cal P}\sum_{i\ne j}\left[\sum_{\sigma}-t_{ij}c^\dagger_{i,\sigma}c_{j,\sigma}+\frac{J_{ij}}{2}\left({\mbox{\boldmath $S$}}_i\cdot\mbox{\boldmath $S$}_j-\frac{n_i n_j}{4}\right)\right]{\cal P}
\label{tjhamiltonian}
\end{equation}
for the one-dimensional chain with even number of site $N$. Here  ${\cal P}$ is the projection operator onto the Fock space without double occupation. $c^\dagger_{i,\sigma}$ ($c_{i,\sigma}$) and $\mbox{\boldmath $S$}_i$ denote the electron creation (annihilation) operator with spin $\sigma(=\uparrow,\downarrow)$ and the spin operator at site $i\in [0,N-1]$, respectively. The transfer energy $t_{ij}$ and exchange energy $J_{ij}$ between the sites $i$ and $j$ are related to each other by the supersymmetric condition: $t_{ij}=J_{ij}/2=D^{-2}_{ij}$. Here $D_{ij}$ is given by $(N/\pi) \sin\left[\pi(i-j)/N\right]$.   

The hole propagator at half-filling is defined by
\begin{equation}
G(x,t)=\langle \phi_{\rm HS}\vert c^\dagger_{x,\sigma}\left(t\right)c_{0,\sigma}(0)\vert \phi_{\rm HS}\rangle/\langle \phi_{\rm HS},\phi_{\rm HS}\rangle, \label{propagator}
\end{equation}
where $\vert \phi_{\rm HS}\rangle $ denotes the singlet ground state at half filling. $c^\dagger_{x,\sigma}(t) $ is the Heisenberg representation of $c^\dagger_{x,\sigma}$. The spatial separation $x$ takes an integer value. We obtain the following expression for $G_{\rm 1s1h}(x,t)$ in the thermodynamic limit:
\begin{equation}
G_{\rm 1s1h}(x,t)=\frac{\left(-1\right)^x}{\pi^3}\int_{0}^\pi {\rm d}q_{\rm s}\int_{0}^{\pi -q_{\rm s}}{\rm d}q_{\rm h}\sqrt{\frac{\pi-q_{\rm s}}{q_{\rm s}}}{\rm exp}\left[-{\rm i}t \left(\epsilon_{\rm h}(q_{\rm h})+\epsilon_{\rm s}\left(q_{\rm s}\right)-\mu\right)\right]\cos\left[\left(q_{\rm h}+q_{\rm s}\right)x\right],
\label{hole-therm}
\end{equation}
where $\epsilon_{\rm h}(q)= (q-\pi/2)^2$ and  $\epsilon_{\rm s}(q)= q(\pi-q)$ denote the dispersion of holon and spinon, respectively. The chemical potential $\mu$ is given by $-\pi^2 /12$ \cite{KY}. The expression (\ref{hole-therm}) is our main result. 

The spectral function  $A(\omega,Q)$ is defined by $G(x,t) =(2\pi)^{-2}\int_{-\infty}^{\infty} {\rm d}\omega \int_{0}^{2\pi} {\rm d}Q{\rm e}^{-{\rm i}\omega t+{\rm i}Q x }A(\omega,Q)$. In  $A(\omega,Q)$, the (1s1h) contribution $A_{\rm 1s1h}(\omega,Q)$ for $Q\in [0,\pi]$ is given by
$$
\frac{\theta\left[\left(\omega'-\epsilon_{\rm h}(Q)\right)\left(\epsilon_{\rm s}(Q)+\pi^2 /4-\omega'\right)\right]}{\pi Q}\sqrt{\frac{\left(Q+\pi/2\right)^2-\omega'}{\omega'-\epsilon_{\rm h}(Q)}}, 
$$
with $\omega'=\omega+\mu$. Here $\theta[x]$ denotes the Heviside step function. The compact support of $A_{\rm 1s1h}(\omega,Q)$ is shown in Fig. 1 by the dark region. In addition, the support of the full spectral function $A(\omega,Q)$ can be derived as the half-filling limit of the Ha-Haldane result \cite{HaHaldane}; the full support is also shown in Fig. 1 as the joint area of dark and light regions. The function $A_{\rm 1s1h}(\omega,Q)$ has the inverse square root singularity along the curve $\omega=\epsilon_{\rm h}(Q)-\mu$ indicated by the bold one. This singularity corresponds to the ``main band'' in the $U\rightarrow\infty$ Hubbard model \cite{SP,Penc2}. In our model, however, there is no equivalent of the ``shadow band'' within the one spinon one holon contribution. 

Next we consider the static limit. For $Q\in [0,\pi]$, the momentum distribution function is given by
$$
G_{\rm 1s1h}(Q,t=0)=\int_{-\infty}^{\infty}{\rm d}x {\rm e}^{-{\rm i}Q x}G_{\rm 1s1h}(x,t=0)=\pi^{-1}{\rm ArcSin}\sqrt{1-Q/\pi}+\pi^{-2}\sqrt{Q\left(\pi-Q\right)}.
$$
In \cite{MetznerVollhardt}, on the other hand, the full result was found to be $G(Q,t=0)=1/2$. In 
Fig. 2, we show $G_{\rm 1s1h}(Q,t=0)$ and $G(Q,t=0)=1/2$. We can see that the (1s1h) contribution is dominant near $Q=0$ while most contribution near $Q=\pi$ comes from (3s1h) states. Additionally, we find that $G_{\rm 1s1h}(x=0,t=0)=3/8$. This means that the (1s1h) contribution amounts to 75\% of the mean density 1/2 with fixed spin. 
 
Now we present the derivation of (\ref{hole-therm}). We denote a state vector $\vert \phi\rangle$ without double occupation by  
\begin{equation}\vert \phi\rangle=\sum_{\left\{x\right\}, \left\{y\right\}} \phi(\left\{x\right\}, \left\{y\right\})\prod_{i\in \left\{x\right\}}S^-_{i} \prod_{j\in \left\{y\right\}}h^\dagger_j \vert F\rangle.  
\end{equation}
Here $\left\{x\right\}=\left\{x_1,\cdots,x_M\right\}$ denotes the set of coordinates for $M$ down-spin electrons and $\left\{y\right\}=\left\{y_1,\cdots,y_P\right\}$ denotes those of $P$ holes. The state $\vert F\rangle$ is the fully-polarized up-spin state. The operators $S^-_i=c^\dagger_{i,\downarrow}c_{i,\uparrow}$ and $h^\dagger_i=c_{i,\uparrow}$ represent the spin-lowering and hole creation operator of site $i$, respectively. 

The wavefunction $\phi\left(\left\{x\right\}, \left\{y\right\}\right) $ is defined only when each argument takes an integer value. However we can regard it as a restriction of a complex function $\Phi \left(\left\{X\right\},\left\{Y\right\}\right)$ of $\left\{X\right\}=\left\{X_1,\cdots,X_M\right\}\in \mbox{\boldmath$C$}^M$ and $\left\{Y\right\}=\left\{Y_1,\cdots,Y_P\right\} \in \mbox{\boldmath$C$}^P$ if the relation $\phi\left(\left\{x\right\},\left\{y\right\}\right)=\Phi\left(\left\{X\right\},\left\{Y\right\}\right)$ hold for $X_k=\exp({\rm i}2\pi x_k/N)$ with $k \in [1,M]$ and $Y_l=\exp({\rm i}2\pi y_l/N)$ with $l \in [1,P]$. We can impose further two conditions on $\Phi \left(\left\{X\right\},\left\{Y\right\}\right)$ without loss of generality: 1) symmetric under pairwise interchange of $X$. 2) antisymmetric under pairwise interchange of $Y$. 3) zero when $X_k=X_l$, $Y_k=Y_l$ or $X_k=Y_l$. 

%
%
 From now on we concentrate on the SU(2) Yangian highest weight states (YHWS)\cite{talstra,note} because the (1s1h) states belong to those states. The YHWS are the states with polynomial-type wavefunctions where the degree of each variable is in the range $(0,N)$. The YHWS form a subset of the Fock space and the residual states can be generated by successive actions of a Yangian generator on YHWS \cite{talstra}. For YHWS, it is known \cite{Wang} that wavefunctions are given by those of the SU(1,1) Calogero-Sutherland model
\begin{eqnarray}
\tilde {\cal H}\Phi&\equiv &\left[\sum_{i=1}^{M+P}\left(Z_i\frac{\partial}{\partial Z_i}-\frac{N}{2}\right)^2-\sum_{i<j}\frac{2Z_i Z_j}{\left(Z_i- Z_j\right)^2}\left(1+s_{ij}\right)\right]\Phi\nonumber\\
&=&\varepsilon \Phi.\label{CS}
\end{eqnarray} 
Here we denote $Z\equiv \left(X,Y\right)\equiv \left(X_1\cdots X_M,Y_1\cdots,Y_P\right)$. $s_{ij}$ is the coordinate exchange operator of the pair $i$ and $j$. The eigenvalue $\varepsilon$ in (\ref{CS}) is related to $E$ in (\ref{tjhamiltonian}) as 
\begin{equation}
E=2\pi^2 \left(\varepsilon+\varepsilon_{M,P}\right)/N^2,
\end{equation}
with $\varepsilon_{M,P} =N\left(N^2 -1\right)/12-(N^2+2)P/4-N^2 M/4$.  
Wavefunctions for YHWS have the form
\begin{equation}
\Phi=\tilde \Phi \prod_{i<j}^M \left(X_i -X_j\right)\prod_{i<j}^{M+P}\left(Z_i -Z_j\right)\prod_{i=1}^{M+P}Z_i. 
\end{equation}
Here $\tilde \Phi(X,Y)$ is a polynomial symmetric with respect to the interchange of $(X_i,X_j)$ and that of $(Y_i,Y_j)$. The degree of $X_i$ in $\tilde\Phi$ is less than or equal to $N_{\rm sp}\equiv N-2M-P$ \cite{Wang}. We can regard $N_{\rm sp}$ as the number of spinon excitations.  The relation $S^z_{\rm tot}=N_{\rm sp}/2$ holds only in YHWS.  

Now suppose that a hole with $S^z =1/2$ is injected to the singlet ground state at half-filling. Then the states with $S_{\rm tot}^z=1/2$ and $P=1$ are relevant intermediate states; $P=1$ means that one holon is excited.  In the intermediate states, only (1s1h) states ($N_{\rm sp}=1$) are YHWS because $2S_{\rm tot}^z=1$. In the following, we diagonalize the Hamiltonian (\ref{CS}) for $M=M_0\equiv N/2-1$ and $P=1$. 

First, we define the basis function $\Psi_{JK}$ for $J\in [0,M_0]$ and $K\in[0,M_0]$ as $\Psi_{JK}=Y^J e_K (X)\Phi_0$. 
Here $e_K (X)$ is the elementary symmetric function of order $K$, which is defined by $\sum_{1\le k_1<k_2\cdots k_K\le M_0} \prod_{i\in \left\{k\right\}}X_i$ with $\left\{k\right\}=\left\{k_1,k_2,\cdots,k_K\right\}$. The function $\Phi_0$ is defined by
\begin{equation}
\Phi_0=Y\prod_{i=1}^{M_0} X_i\prod_{i<j}^{M_0}\left(X_i -X_j\right)^2\prod_{i=1}^{M_0}\left(X_i -Y\right).
\end{equation}

Now we consider the case $J+K<M_0$. In the present basis, the diagonal element $\langle \Psi_{JK}\vert \tilde {\cal H}\vert \Psi_{JK}\rangle$ is given by
$\varepsilon_{JK}=2J^2-2JM_0-2K^2+2KM_0+J+K-M_0+\varepsilon_{\rm HS}\label{ejk1}$ 
with $\varepsilon_{\rm HS}=(N^2/4-1)N/6$. The off-diagonal element $\langle \Psi_{J+L,K-L}\vert \tilde {\cal H}\vert \Psi_{J,K}\rangle$ is $(-1)^L 2(M_0-J-K)$ for  $L>0$ and zero for $L<0$. The Hamiltonian matrix in this basis is triangular and hence it can be solved recursively. As a consequence, we obtain 
\begin{equation}
\Phi_{JK}=\sum_{L=0}^{K}a_L \Psi_{J+L,K-L}\label{jk1},  
\end{equation}
for $J+K<M_0$. Here the coefficients $a_L$ are given by $a_0=1$ and
$a_L=(-1)^{L+1}2^{-1}\pi^{-1/2}\Gamma\left[L-1/2\right]/L!$ for $L\ge 1$.
For $J+K\ge M_0$, the eigenfunction is obtained similarly as 
\begin{equation}
\Phi_{JK}=\sum_{L=0}^{M_0-K}a_L \Psi_{J-L,K+L}.\label{jk2}
\end{equation}
with the eigenvalue $\varepsilon_{JK}=2J^2-2JM_0-2K^2+2KM_0+M_0-J-K+\varepsilon_{\rm HS}\label{ejk2}$.

With these preliminaries, we calculate the (1s1h) contribution to the hole propagator 
\begin{eqnarray}
& &G_{\rm 1s1h}(x,t)\nonumber\\
&=&\sum_{J=0}^{M_0} \sum_{K=0}^{M_0} \vert\langle \Phi_{JK}\vert
c_{0,\downarrow}\vert \Phi_{\rm HS}\rangle\vert^2/\left(\langle \Phi_{JK}|\Phi_{JK}\rangle\langle \Phi_{\rm HS}|\Phi_{\rm HS}\rangle\right)\nonumber\\
& &(-1)^x\exp\left[-{\rm i}t \left(E_{JK}-E_{\rm HS}\right)+{\rm i}2\pi\left(J+K+1\right)x/N\right]\nonumber\label{G1s1h}. 
\end{eqnarray}
Here the ground state wavefunction $\Phi_{\rm HS}$ at half-filling is given by  \cite{Haldane,Shastry}
\begin{equation}
\Phi_{\rm HS}=\prod_{i<j}^{N/2}\left(X_i -X_j\right)^2\prod_{i=1}^{N/2}X_i \label{hs}. 
\end{equation}
The eigenvalues $E_{JK}$ and $E_{\rm HS}$ of $\vert\Phi_{JK}\rangle$ and  $\vert\Phi_{\rm HS}\rangle$ for ${\cal H}_{\it tJ}$ in (\ref{tjhamiltonian}) are given by $2\pi ^2 \left(\varepsilon_{JK}+\varepsilon_{M_0,1}\right)/N^2$ and $2\pi ^2 \left(\varepsilon_{\rm HS}+\varepsilon_{M_0+1,0}\right)/N^2$, respectively.   
In the following, we consider the norm first and then discuss the matrix element. 

For general $M$ and $P$, the inner product between $\vert \Phi\rangle$ and $\vert \Phi'\rangle$ can be written as
\begin{equation}
\langle \Phi,\Phi'\rangle=N^{M+P}\left[\Phi,\Phi'\right]_0/(M!P!), \label{constant}
\end{equation} 
 when the wavefunctions $\Phi(Z)$, $\Phi'(Z)$ are polynomials of $Z$. Here we define a scalar product $\left[\Phi,\Phi'\right]_0$ on polynomials by the constant term in $ \Phi(1/Z)\Phi'(Z)$; the symbol $1/Z$ denotes $\left(1/Z_1,\cdots,1/Z_{M+P}\right)$. 

The norm $[\Phi_{\rm HS},\Phi_{\rm HS}]_0$ for (\ref{hs}) has been known as $A_N=2^{-N/2}N!$ \cite{Ha,Lesage}. 
%
%
In addition, the norm $[\Phi_{JK},\Phi_{JK}]_0$ for (\ref{jk1}) or (\ref{jk2}) is obtained as
\begin{equation}
\frac{A_N}{N}\frac{\Gamma\left[N/2-K-\Delta_{JK}\right]\Gamma\left[K+1/2+\Delta_{JK}\right]}{\Gamma\left[N/2-K+1/2-\Delta_{JK}\right]\Gamma\left[K+1+\Delta_{JK}\right]}\label{JKintegralnorm},  
\end{equation} 
from the theory of  a variant of non-symmetric Jack polynomials\cite{Dunkl}. In (\ref{JKintegralnorm}), $\Delta_{KJ}$ is 1/2 for $J+K\ge M_0$ or zero otherwise.

Next we consider the matrix element $\langle \Phi_{JK}\vert c_{0,\downarrow} \vert \Phi_{\rm HS}\rangle=\langle \Phi_{JK}\vert h^\dagger_0 S^+_0 \vert \Phi_{\rm HS}\rangle$. We present the derivation only for $J+K<M_0$ because the following discussions are applicable also to the case $J+K\ge M_0$.  

Now we introduce homogeneous symmetric polynomials $J_\mu(X)=J_{\mu_1,\mu_2}(X)$ of $X=\left\{X_1,\cdots,X_{M_0}\right\}$ with $\mu=\left(\mu_1,\mu_2\right)$ by 
\begin{equation}
e_K(X) e_{K'}(X)=\sum_{\mu_1=K}^{M_0}\sum_{\mu_2=0}^K\delta_{K+K',\mu_1+\mu_2}d_\mu^K J_\mu(X) \label{pieri} 
\end{equation}
for $K,K'\in [0,M_0]$. 
Here $\delta_{a,b}$ denotes the Kronecker's delta and the coefficient $d_\mu^K$ is given by
$$
\frac{\Gamma\left[\mu_1 -K +1/2\right]\Gamma\left[K-\mu_2+1/2\right]\Gamma\left[\mu_1 -\mu_2 +1\right]}{\Gamma\left[1/2\right]\Gamma\left[\mu_1 -K +1\right]\Gamma\left[K-\mu_2+1\right]\Gamma\left[\mu_1 -\mu_2 +1/2\right]}. 
$$
The polynomials $J_\mu$ are called the symmetric Jack polynomials \cite{Macd}. They are useful in our calculation because they form an orthogonal basis with the following norm \cite{Macd}:
\begin{eqnarray}
[J_\mu \tilde \Phi_{\rm HS},J_\nu \tilde \Phi_{\rm HS}]_0
&=&\frac{\delta_{\mu,\nu}N!}{2^{N/2}\pi}\frac{\Gamma\left[\mu_1 -\mu_2 +1/2\right]\Gamma\left[\mu_1-\mu_2 +3/2\right]}{\Gamma^2\left[\mu_1 -\mu_2 +1\right]}\nonumber\\
& &\prod_{i=1}^2 \frac{\Gamma\left[\mu_i+(3-i)/2\right]\Gamma\left[M_0-\mu_i+i/2\right]}{\Gamma\left[\mu_i+2-i/2\right]\Gamma\left[M_0-\mu_i+(i+1)/2\right]}. \label{jack} 
\end{eqnarray}
Here 
$
\tilde \Phi_{\rm HS}$ denotes $\prod_{i<j}^{M_0}\left(X_i -X_j\right)^2\prod_{i=1}^{M_0}X_i
$. 

The wavefunction of $h_0\vert \Phi_{JK}\rangle$ is given by $\Phi_{JK}(\left\{X_i\right\}_{i=1}^{M_0},Y=1)$. With the use of (\ref{pieri}), this wavefunction can be written as
\begin{equation}
\sum_{L=0}^K \sum_{\mu_1=K-L}^{M_0}\sum_{\mu_2=0}^{K-L}\left(-1\right)^{L-\mu_1-\mu_2}a_L d_\mu^{K-L} J_\mu \tilde \Phi_{\rm HS},   
\label{decomp1}
\end{equation}
up to a phase factor.

On the other hand, the wavefunction of $S^+_0 \vert \Phi_{\rm HS}\rangle$ is given by \cite{HZ} $\prod_{i=1}^{M_0}\left(X_i -1\right)^2\tilde \Phi_{\rm HS}\label{HZ}$. Further the decomposition of this wavefunction in terms of $J_\mu \tilde \Phi_{\rm HS}$ is known as\cite{Ha,Lesage} 
\begin{equation}
\sum_{\mu_1\ge\mu_2}(-1)^{\mu_1 +\mu_2}B_{\mu_1,\mu_2}J_{\mu_1,\mu_2}\tilde \Phi_{\rm HS}, \label{decomp}
\end{equation}
with $B_{\mu_1,\mu_2}=\Gamma\left[1/2\right]\Gamma\left[\mu_1-\mu_2+1\right]/\Gamma\left[\mu_1-\mu_2 +1/2\right]$. 
From the expressions (\ref{constant}), (\ref{jack}), (\ref{decomp1}) and (\ref{decomp}), we obtain a simple expression for the matrix element
\begin{equation}
\vert\langle \Phi_{JK}\vert h^\dagger_0 S^+_0 \vert \Phi_{\rm HS}\rangle\vert=g_N \Gamma\left[K+1/2\right]/\Gamma\left[K+1\right]
\label{fk1}
\end{equation}
%
%
with  $g_N=\pi^{-1/2}(N/2)^{N/2-1} (N-1)!/(N/2-1)!$. 

A similar calculation for $J+K\ge M_0$ gives
\begin{equation}
\vert\langle\Phi_{JK}\vert h^\dagger_0S^+_0\vert\Phi_{\rm HS}\rangle\vert=g_N \Gamma[M_0-K-1/2]/\Gamma[M_0-K]. \label{fk2}
\end{equation}
As a result of (\ref{JKintegralnorm}), (\ref{fk1}) and (\ref{fk2}), we obtain the expression for the hole propagator
\begin{eqnarray}
& &G_{\rm 1s1h}(x,t)=1/(2N)\nonumber\\
&+&\frac{4\left(-1\right)^x}{\pi N^2}\sum_{K=0}^{N/2-2} \sum_{J=0}^{N/2-K-2}
\frac{\Gamma\left[N/2-K+1/2\right]\Gamma\left[K+1/2\right]}{\Gamma\left[N/2-K\right]\Gamma\left[K+1\right]}\nonumber\\
& &{\rm e}^{-{\rm i}t E'_{JK}}{\rm cos}\left({\rm i}2\pi x(J+K+1)/N\right), 
\label{hole-finite}
\end{eqnarray}
with $E'_{JK}=2\pi^2 \left(2J^2-2JM_0-2K^2+2KM_0+J+K-M_0-1/6\right)/N^2+\pi^2 /3$. Further we introduce $q_{\rm h}=2\pi J/N$  and $q_{\rm s}=2\pi K/N$ and take the thermodynamic limit $N\rightarrow \infty$. Finally, the expression (\ref{hole-finite}) reduces to (\ref{hole-therm}). 
 
I thank J. Zittartz for his hospitality. I also thank Y. Kuramoto for bringing my attention to this problem, and Y. Saiga and M. Arikawa for informative discussions and pointing out an error in the original manuscript. This work was supported by Sonderforschungsbereich 341 K\"oln-Aachen-J\"ulich. 

\begin{figure}
\caption{Region of non-vanishing spectral weight function of the hole propagator for $Q\in [0,\pi]$. The dark region represents the contribution from states with one spinon and one holon. At the lower edge indicated by the bold line, the spectral weight has the inverse square root singularity. } 

\caption{The contribution from states with one spinon and one holon to the momentum distribution function $G_{\rm 1s1h}(Q,t=0)$ for $Q\in [0,\pi]$. Full result $G(Q,t=0)=1/2$ is also shown. }
\end{figure}

\begin{references}
%
%
\bibitem[*]{permanent}
On leave from Department of Applied Physics, 
University of Tokyo, Tokyo 113-8656, Japan. 
%
\bibitem{Kim}C. Kim {\it et al.}, Phys. Rev. Lett. {\bf 77}, 4054 (1996).
%
%
%
%
%
%
\bibitem{SP}S. Sorella and A. Parola, J. Phys. Cond. Mat. {\bf 4}, 3589 (1992);
A. Parola and S. Sorella, Phys. Rev. B {\bf 45}, 13156 (1992)
%
%
\bibitem{Penc2}K. Penc {\it et al.}, Phys. Rev. Lett. {\bf 77}, 1390 (1997); J. Favand {\it et al.}, Phys. Rev. B  {\bf 55}, R4859 (1997)
\bibitem{KY}Y. Kuramoto and H. Yokoyama, Phys. Rev. Lett. {\bf 67}, 1338 (1991).
\bibitem{KatoKuramoto}Y. Kato and Y. Kuramoto, J. Phys. Soc. Jpn. {\bf 65}, 1622 (1996). 
\bibitem{Haldanefrac}
F. D. M. Haldane, 
Phys. Rev. Lett. {\bf 67}, 937 (1991).  

\bibitem{Haldane}F. D. M. Haldane, Phys. Rev. Lett. {\bf 60}, 635 (1988). 
\bibitem{Shastry}B. S. Shastry, Phys. Rev. Lett. {\bf 60}, 639 (1988). 
\bibitem{HZ}F. D. M. Haldane and M. R. Zirnbauer, Phys. Rev. Lett. {\bf 71}, 4055 (1993). 
\bibitem{HaHaldane}Z. N. C. Ha and F. D. M. Haldane, Phys. Rev. Lett. {\bf 73}, 2887 (1994).
\bibitem{Cal}
F.~Calogero,
J. Math. Phys. {\bf 10}, 2191, 2197 (1969).

\bibitem{Suth}
B.~Sutherland,
Phys. Rev. A{\bf 4}, 2019 (1971),
A{\bf 5} 1372 (1972),
J. Math. Phys. {\bf 12}, 246, 251 (1971).


\bibitem{Opdam}
E. Opdam,
Acta. Math. {\bf 175}, 75 (1995).

\bibitem{Sahi}
S. Sahi,
IMRN {\bf 20}, 997 (1996); F. Knop and S. Sahi, Inv. Math. {\bf 128}, 9 (1997).


\bibitem{BF1}
T. H. Baker and P. J. Forrester,
Nucl. Phys. B {\bf 492}, 682 (1997); preprint (1997) q-alg/9707001.

\bibitem{Dunkl}
C. F. Dunkl, preprint (1997) q-alg/9710015, To be published in Commun. Math. Physics.

\bibitem{MetznerVollhardt}T. Kaplan, P. Horsch and P. Fulde, Phys. Rev. Lett. {\bf 49}, 889 (1982); W. Metzner and D. Vollhardt, Phys. Rev. B, {\bf 37}, 7383 (1988). 
\bibitem{talstra}See. e.g. J. C. Talstra, PhD Thesis (Princeton University 1995) and references therein. 
\bibitem{note} We fix the number of holes. Hence the residual quantum symmetry is Y(su2). Our method is a variant of that used in \cite{Wang}.
\bibitem{Wang}D. F. Wang, J. T. Liu and P. Coleman, Phys. Rev. B. {\bf 46}, 6639 (1992).
 
%
%


\bibitem{Ha}
Z.~N.~C.~Ha,
Phys. Rev. Lett. {\bf 73}, 1574 (1994),
{\bf 74}, 620 (1995) (errata),
Nucl. Phys. B{\bf 435}, 604 (1995).

\bibitem{Lesage}
F.~Lesage, V.~Pasquier and D.~Serban,
Nucl. Phys. B{\bf 435}, 585 (1995).



\bibitem{Macd}

I.~G.~Macdonald,
{\it Symmetric Functions and Hall Polynomials, 2nd ed.},
(Oxford University Press, 1995).  $J_{\mu_1,\mu_2}$ in this paper corresponds to $P_\lambda ^{(\alpha=1/2)}$ with $\lambda'=\left\{\mu_1,\mu_2\right\}$ in section 10 of chapter VI of Macdonald's book.  

\end{references}
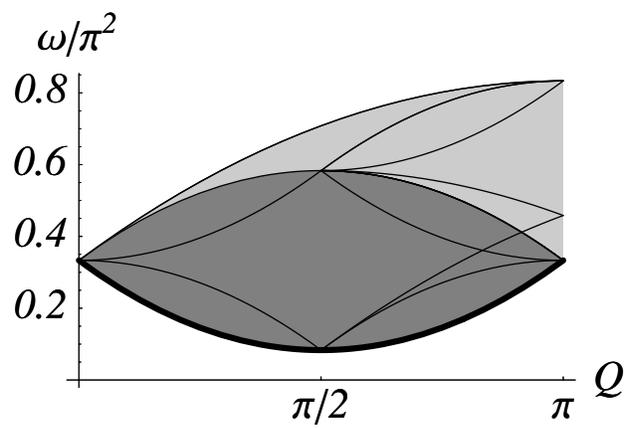
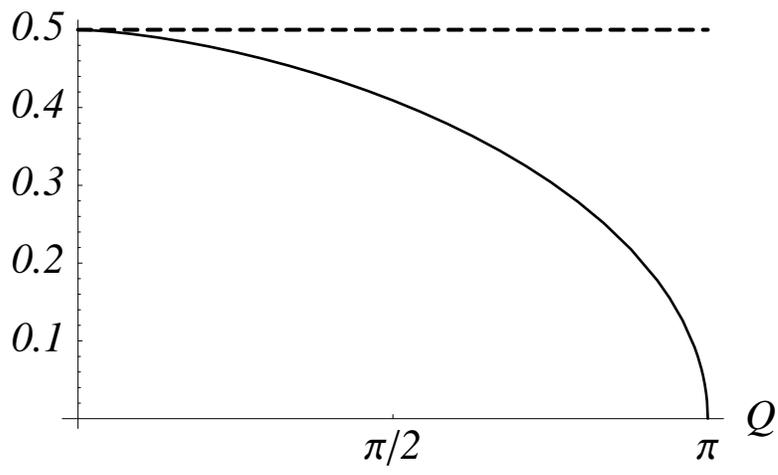
\end{document}